\documentclass[12pt]{iopart}
\usepackage{graphicx}
\usepackage{multirow}
\usepackage{makecell}
\usepackage{xcolor}
\newcommand{\gt}{>}
\begin{document}
\title[Bridging the Synthesizability Gap in Perovskites: A PU-Learning Framework]{Bridging the Synthesizability Gap in Perovskites by Combining Computations, Literature Data, and PU Learning}
\author{Rushik Desai$^1$, Junyeong Ahn$^1$, Alejandro Strachan$^1$,
Arun Mannodi-Kanakkithodi$^1$}
\address{$^1$ School of Materials Engineering, Purdue University, West Lafayette,
IN 47907, USA.}
\ead{amannodi@purdue.edu}
\begin{abstract}
Among emerging energy materials, halide and chalcogenide perovskites have garnered significant attention over the last decade owing to the abundance of their constituent species, low manufacturing costs, and their highly tunable composition-structure-property space. Navigating the vast perovskite compositional landscape is possible using density functional theory (DFT) computations, but they are not easily extended to predictions of the synthesizability of new materials and their properties. As a result, only a limited number of compositions identified to have desirable optoelectronic properties from these calculations have been realized experimentally. One way to bridge this gap is by learning from the experimental literature about how the perovskite composition-structure space relates to their likelihood of laboratory synthesis. Here, we present our efforts in combining high-throughput DFT data with experimental labels collected from the literature to train classifier models employing various materials descriptors to forecast the synthesizability of any given perovskite compound. Our framework utilizes the positive and unlabeled (PU) learning strategy and makes probabilistic estimates of the synthesis likelihood based on DFT-computed energies and the prior existence of similar synthesized compounds. Our data and models can be readily accessed via a Findable, Accessible, Interoperable, and Reproducible (FAIR) nanoHUB tool.
\end{abstract}

\maketitle

\section{Introduction}

Perovskites and their derivatives have become increasingly prominent for a variety of applications over the last couple of decades. Due to their tunable band gaps (0.5 eV to 6 eV) \cite{bandgap-1, bandgap-2}, high light absorption coefficients ($\gt10^5$ cm$^{-1}$) \cite{absorption-coeff}, and low cost of synthesis, they are highly sought-after commercial replacements for Si as solar absorbers. ABX$_3$ halide perovskites (HaPs) have been the most studied subclass of materials for photovoltaics \cite{halide-photovol} and photocatalysis \cite{halide-photocat}. Among them, hybrid organic-inorganic perovskites, which have an organic cation such as methylammonium or formamidinium at the A-site, have shown the best photovoltaic efficiencies \cite{hybrid-1}, but have also suffered from thermal and mechanical instability \cite{hybrid-instability-1, hybrid-instability-2}. Hybrid perovskites can undergo protonation in the presence of moisture and exhibit undesired phase changes at elevated temperatures \cite{hybrid-instability-thermal}. Inorganic perovskites do not have the same issues but typically contain Pb at the B-site and thus suffer from Pb toxicity \cite{lead-toxicity-1}. \\

The many challenges of halide perovskites are being actively tackled via composition engineering \cite{lead-free-1, composition-tuning-1, composition-tuning-2, stability}, including complex alloying at anion and/or cation sites \cite{alloying-1, alloying-2}. Much effort has recently been put into going beyond halide perovskites to chalcogenide perovskites (ChPs) \cite{chalco-perovs, chalco-perovs-2} and various perovskite derivatives \cite{derivative-1, derivative-2}. Compounds in many perovskite-derived chemical classes, such as A$_2$BB'X$_6$ double perovskites (DP), A$_2$BX$_6$ vacancy-ordered double perovskites (VO), and A$_3$B$_2$X$_9$ compounds, effectively overcome some of the aforementioned issues. Due to the presence of such different types of perovskites, the exploration space is massive and complex. High-throughput density functional theory (HT-DFT) computations have been extensively used to perform screening across these chemical spaces \cite{ht-dft-abse3, ht-dft-vo, ht-dft-dp} since it is infeasible to synthesize and characterize all possible compounds without prior prediction of their properties. DFT datasets help researchers comprehensively map the composition-structure-property space of perovskites to aid experimental studies. \\

We have performed various perovskite screening studies in our research group over the past few years \cite{htdft-1, htdft-2, htdft-3}, leading to a large DFT dataset of ABX$_3$ halide perovskites in a chemical space spanning 14 different A/B/X species and four prototype perovskite phases. This dataset includes inorganic and hybrid perovskites and structures with alloying at any of the cation or anion sites. It includes $>$ 1000 unique perovskite compositions with properties such as band gap, decomposition energy, and spectroscopic limited maximum Efficiency (SLME) \cite{slme} computed from various semi-local GGA and non-local HSE06 functionals \cite{pbe, pbesol, d3-1, d3-2, hse-1, hse-2}. The functionals we used include GGA-PBE with van der Waals D3 corrections \cite{d3-1, d3-2} for hybrid perovskites, GGA-PBEsol for inorganic compounds, and both static and full relaxation HSE06 with and without spin orbit coupling (SOC, important for systems containing heavier atoms). Our recent work \cite{htdft-2} highlights how different functionals affect the properties of interest for different types of perovskites, providing some understanding about the levels of theory necessary for reproducing experimental results. While a majority of our previous dataset is focused on halide perovskites, it has now been extended to chalcogenide perovskites and a variety of double perovskites. \\

%These calculations provide an accurate way of mapping structure and property, but are often expensive and slow. Not all compositions have suitable properties, and data generation utilizes many resources. 
Given that our dataset is well-representative of the perovskite chemical space, it is suitable for training surrogate machine learning (ML) models \cite{ml-review, ml-review-2} which can then be deployed for accelerated prediction and screening across millions of possible compositions. Using the DFT data, we trained regularized greedy forest models to predict the band gap and decomposition energy (which shows whether a compound would readily decompose to alternative binary phases) of any ABX$_3$ halide perovskite. This is different from energy above hull that takes into account all possible combinations of phases (theoretical and experimental). We choose decomposition energy as a possible stability metric because the primary decomposition pathway for inorganic and hybrid perovskites is to their binary phases \cite{decompose}, and most perovskite films are synthesized using these known binary precursors \cite{binary}. These properties are central in determining the stability and photovoltaic suitability of the materials. Since we have data from different levels of theory, we trained these models within a multi-fidelity framework \cite{ml-amk-2} such that the source of data (or DFT functional) is part of the input. We used both compositional information and well-known elemental or molecular properties as part of the descriptors and achieved prediction accuracy of $\sim$ 98\% for the properties of interest. We used K-fold cross-validation to evaluate model overfitting. \textbf{Figure S1} plots the training and testing RMSE values for decomposition energy and band gap across different folds, showing a consistency in prediction and generally no overfitting.\\

The best regression models were used to make on-demand predictions for thousands of enumerated ABX$_3$ compounds across the four prototype phases, leading to the discovery of (a) novel stable high-efficiency single-junction solar absorber candidates \cite{ml-amk-2}, and (b) novel candidates with high predicted solar-to-hydrogen (STH) efficiencies for photocatalytic water splitting \cite{ml-amk-1}. The latter work utilized empirically predicted electronic band edges to place them relative to H$_2$O redox potential levels. Importantly, the best materials found via DFT+ML screening were also divided into different categories, such as hybrid vs purely inorganic, and Pb-free vs Pb-containing compounds. We recently extended this work to structure-based predictions using crystal graph neural network models \cite{gnn-1, gnn-2, gnn-3}. By accumulating a massive perovskite dataset of $\sim$ 30,000 structures and their DFT-computed energies, atomic forces, and stresses from across all our recent publications, we ``fine-tuned" the M3GNET model \cite{m3gnet} to develop a comprehensive interatomic potential that is an effective DFT surrogate for our chosen perovskite chemical space \cite{perovsiap}. Our rigorously optimized ``Perovs-IAP" model makes accurate energy and force predictions for bulk, surface slab, and defect-containing perovskite structures, and is capable of rapid geometry optimization of any such structure. \\

Thus, we now have a rich perovskite dataset and versatile descriptor-based and structure-based predictive models. However, DFT predictions of the decomposition energy or energy above hull are not entirely indicative of the likelihood of synthesis of any novel compound. In the past, the Goldschmidt tolerance factor \cite{gs} and octahedral factor \cite{octahedra} have been used to determine the general suitability of constituent ions to form a structurally robust perovskite based on their ionic radii, but satisfying these factors does not necessarily translate to a compound being synthesizable, especially when it comes to complex perovskite alloys with a distribution of anions and cations. DFT energies are often used for evaluating formability, but are clearly insufficient because calculations are performed at 0 K and do not account for entropic stabilization and disorder unless specifically included. What comes to the rescue here is the existence of many perovskite compounds that were actually synthesized and reported in the literature, and may inform the synthesis likelihood of other similar materials identified via computations. \\

%In our dataset, we apply decomposition energies, which quantify the propensity of the perovskite to decompose into its binaries. This approach again compares energies from 0 K calculations and is not entirely deterministic. \\

Classification techniques could be used for synthesis predictions provided the availability of sufficient positive and negative labels, but since studies only report positive examples, typical classifiers are not applicable. As material properties are emergent, if we can map their chemical descriptors to a synthesis label, we could, in theory, predict the order of synthesizability of different compositions. However, this requires a semi-supervised learning approach since it may be nearly impossible to classify a material as definitely non-synthesizable. Publications seldom report compositions that were attempted but failed to materialize in the lab. This means one could only apply positive labels to some of the materials and unknown labels to the remaining materials, which have not been synthesized to date but could be in the future. One technique that has recently been used by researchers for synthesis predictions is ``Positive-Unlabeled (PU) learning" \cite{pu-learning, syncotrain, synthesizability-review}, where the classifier learns from the positive examples and assigns a probability to the unlabeled points based on similarity of the descriptors. PU-learning has been theoretically validated \cite{pu-theoretical-1, pu-theoretical-2} for its capacity to extract information from an unlabeled set, even under the constraint of accessing only positive-labeled examples and no negative labels. \\

Frey et al. \cite{mxene} demonstrated the use of PU learning in predicting synthesis scores for MXenes, a novel class of layered materials. They used a decision tree classifier with specific MAX (the bulk phase from which MXenes are derived) and MXene features such as the number of layers, in-plane lattice constant, bond lengths, etc., which were combined with DFT-computed features such as formation energies and Bader charges. The labels for the synthesized compounds were obtained from their work on synthesizing MXenes and MAX phases. The model was able to predict 18 new MXene compositions as synthesizable out of 118 unlabeled samples. Another study by Jang et al. \cite{pucgcnn} used the PU learning approach but combined it with atomic features and the crystal-graph convolutional neural network (CGCNN) framework \cite{cgcnn}. This work used the entire Materials Project (MP) \cite{mp} dataset and provided positive synthesis labels to structures having an Inorganic Crystal Structure Database (ICSD) \cite{icsd} tag. Synthesis probabilities were predicted for materials in the Open Quantum Materials Database (OQMD) \cite{oqmd} and the MP-Virtual database; $\sim$ 12\% of the total compounds from OQMD and $\sim$ 20\% from MP-virtual datasets were found to be synthesizable. \\

Another study on a similar dataset was performed by Gu et al. \cite{pu-perovs} who trained a graph convolution network based on MEGNet \cite{megnet} using a combination of perovskite data from AFLOW \cite{aflow}, Materials Project, and OQMD \cite{oqmd}. Domain-specific transfer learning was performed to make the model accurate for perovskite structures. A recent study by Zhao et al. \cite{mof-pu} utilized a similar approach for Metal Organic Frameworks (MOFs) and used a CGCNN-based classifier model. Compounds were divided based on pre-defined rules, following which their synthesizability scores were predicted. $\sim$ 12\% of the compounds from across the unlabeled set were found to be synthesizable. Yet another recent study from Chung et al. \cite{pu-walsh} predicted the synthesizability of ternary oxides using PU learning, with labels applied specifically for oxides that can be synthesized via solid-state methods. \\

\begin{figure}
    \centering
    \includegraphics[width=0.65\linewidth]{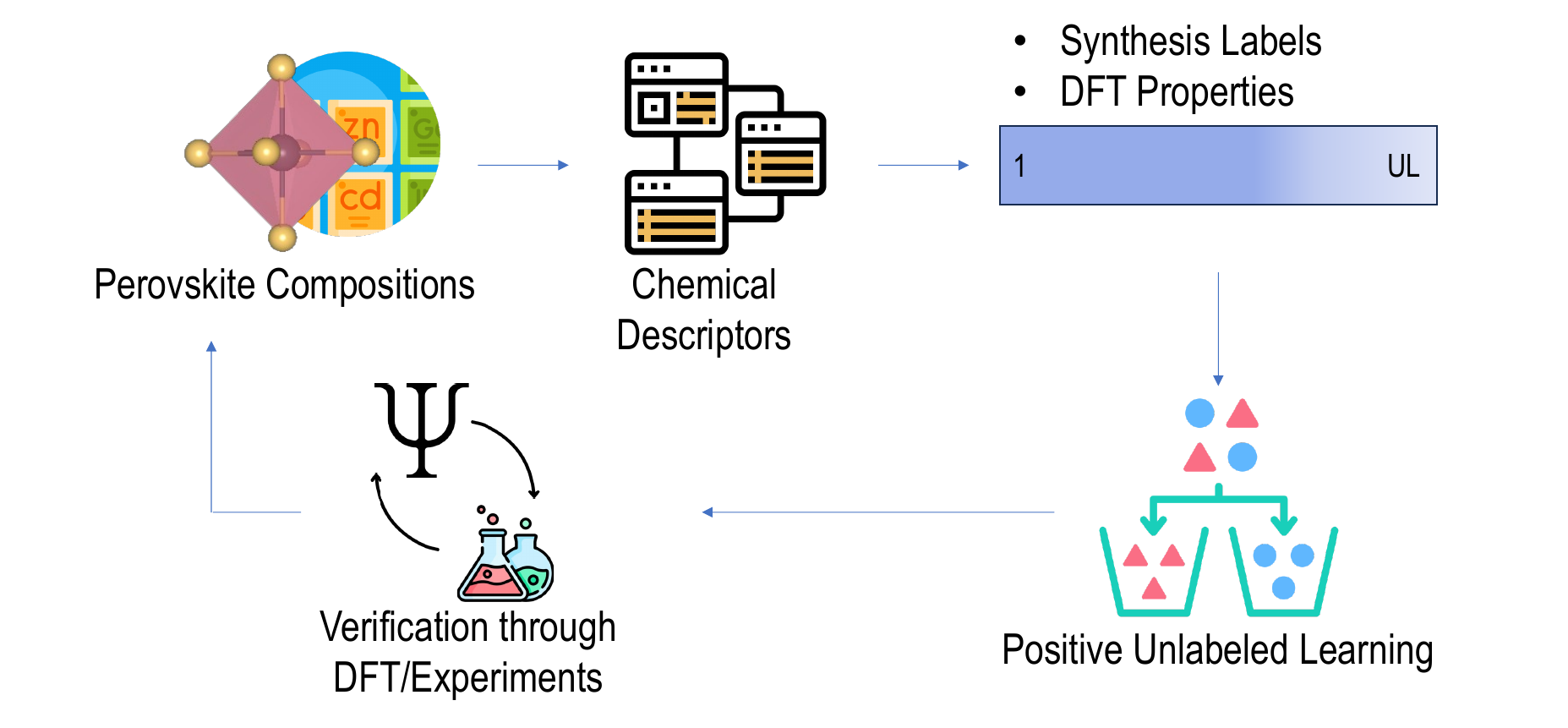}
    \caption{The overall workflow used for designing novel perovskites with high synthesis likelihood and desired DFT-accuracy properties.}
    \label{fig:workflow}
\end{figure}
% \begin{itemize}
%     \item Discuss importance of perovskites for energy applications
%     \item Discuss HT-DFT studies done by the group
%     \item Problem with synthesizing compositions out of ML models and discuss synthesizability predictions
% \end{itemize}
% \clearpage

Many of these studies included structural information of the materials, but this requires ground-state structures to be available for all new materials. It is often more convenient to perform high-throughput screening using compositional information and template structures, before crystal structure-based prediction and understanding is obtained. Thus, in this work, we combined our descriptor-based models described above with the PU learning approach to obtain the probability of synthesis of any A-B-X halide or chalcogenide compound, in addition to its (DFT accuracy) decomposition energy and band gap. Our descriptors utilize well-known elemental or molecular properties of A/B/X ions, such as the electronegativity, ionization energy, electron affinity,  and boiling point, and in addition, one-hot encodes the composition and phase. We further added the DFT properties (band gap and decomposition energy) as descriptors for every material when training a classifier model to predict the synthesis likelihood. \\ 

A major consideration is, of course, the collection of synthesis labels for as many compounds as possible. To do this, we used large language models (LLMs) \cite{Llama32} and extracted chemical formulas of ABX$_3$, A$_2$BX$_6$, or A$_2$BB'X$_6$ halide or chalcogenide compounds reported to be synthesized in various publications. We additionally applied labels to compounds based on their existence in the ICSD, ultimately resulting in 239 compounds with positive synthesis labels. This was used along with a dataset of 832 unlabeled compounds simulated using the GGA-PBE functional to train combined models for predicting both DFT properties and the synthesis probability. Based on prediction and screening across thousands of hypothetical compounds, we discovered potentially hundreds of new stable and synthesizable perovskites with promising optoelectronic properties. Our general workflow is presented in  \textbf{Figure \ref{fig:workflow}}. In the following sections, we describe the computational data generation, experimental label extraction, descriptors, different ML models, and results of training and prediction over unlabeled samples. We discuss the best candidates thus identified and present an outlook for the future of perovskite discovery. \\

\section{Methods}

\subsection{DFT Database}

\subsubsection{Mannodi Group Dataset}

\begin{figure}
    \centering
    \includegraphics[width=\linewidth]{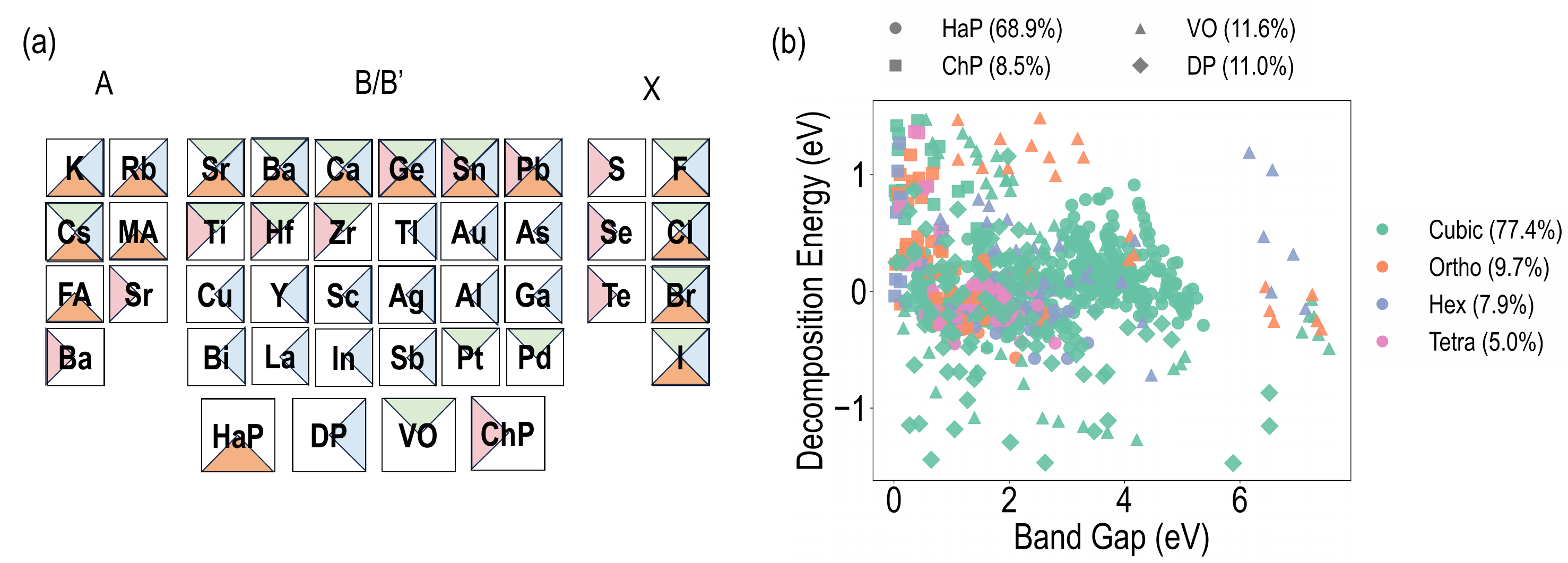}
    \caption{(a) The compositional space considered across the four types of perovskite compounds in our dataset. (b) Visualization of the DFT dataset as a decomposition energy vs band gap plot, with different symbols and colors respectively showing different perovskite types and phases. Here, HaP = ABX$_3$ halide perovskite, ChP = ABX$_3$ chalcogenide perovskite, VO = A$_2$BX$_6$ vacancy-ordered double perovskite, and DP = A$_2$BB'X$_6$ double perovskite.}
    \label{fig:db}
\end{figure}

DFT computations performed in our group over the years and available as part of multiple publications \cite{htdft-1, htdft-2, htdft-3} have resulted in an extensive database of halide and chalcogenide (single and double) perovskite structures across cubic, tetragonal, orthorhombic, and hexagonal phases. All DFT computations were performed using the Vienna Ab-initio Simulation Package (VASP) \cite{DFT1, DFT2}, employing the projector augmented wave (PAW) pseudopotentials \cite{paw1, paw2}. A plane-wave kinetic energy cutoff of 500 eV and a Monkhorst-Pack \textit{k}-mesh of density $\sim$0.03 (2$\pi$Å$^{-1}$) to sample the Brillouin zone was used across all the calculations. A distribution of the ions that adopt the A, B/B', and X sites in these compounds is pictured in \textbf{Figure \ref{fig:db}(a)}, with different colors referring to ABX$_3$ halide perovskites (HaP), ABX$_3$ chalcogenide perovskites (ChP), A$_2$BB'X$_6$ double perovskites (DP), and A$_2$BX$_6$ vacancy-ordered double perovskites (VO). As explained later, the identity and fractions of these A/B/B'/X species, along with their phases, are used to create perovskite feature vectors that are inputs to various ML models. A combinatorially massive number of A-B-X compositions could be created within the highlighted chemical space with arbitrary mixing at any site, and only a small fraction of it was ultimately used for creating the DFT dataset. \\

%We classify the elements as such to obtain site-specific features. This would further help interpret the model using feature importance values. We have included all the elements that have been found in a perovskite matrix so that when we later generate compositions, our models can generalize better. \\ 
\begin{equation}
    \Delta H = E_{ABX_3} - \sum_i x_i E_{AX}
    - \sum_i x_i E_{BX_2} + k_B T \left( \sum_i x_i \ln(x_i) \right)
    \label{eq:decomp}
\end{equation}

\begin{equation}
\begin{array}{l}
    \Delta H\left[AB_{1-x}B'_xX_3\right] = E_{ABX_3} - E_{AX} - x \times E_{BX_2} - (1-x) \times E_{B'X_2} \\
    \qquad\qquad + k_B T \left(x \ln(x) + (1-x) \ln(1-x)\right)
\end{array}
\label{eq:decomp_alloy}
\end{equation}
\textbf{Figure \ref{fig:db} (b)} shows the computed decomposition energy (in eV per ABX$_3$ formula unit) plotted against the band gap (in eV) for a total of 851 compounds, from GGA-PBE calculations. While our prior work also reports data from other GGA and HSE06 functionals, we focus on the PBE dataset for this work because of its relatively larger coverage of the perovskite chemical space of interest. Moreover, the decomposition energy and band gap exhibit a strong correlation between the two functionals, as demonstrated in \textbf{Figure S4}. This suggests that synthesizability predictions derived from either functional should yield consistent trends. Even though the majority of the dataset is cubic and HaP compositions, the models we report later perform well for all phases and ChP, VO, and DP compounds. The band gap values range from close to 0 eV to greater than 6 eV, and decomposition energies range from -1.5 eV (indicating high stability) to 1.5 eV. The formula to calculate the decomposition energy is shown in \textbf{Eq. \ref{eq:decomp}}, where $E_{comp}$ stands for the total energy from DFT for the given composition, and the final term accounts for the mixing entropy in alloys. The entropy term helps account for the stabilization induced by the ionic disorder in the perovskite alloys. Although other entropic contributions are ignored here, using the decomposition energy with mixing entropy information as an input descriptor in synthesizability classification helps encode some entropy information into the model. As an example, the decomposition energy for a compound with B-site mixing will be calculated as shown in \textbf{Eq. \ref{eq:decomp_alloy}}. These formulas hold for both ABX$_3$ HaP and ChP compounds. For A$_2$BB'X$_6$ double perovskites, we use binaries based on both B and B' cations. For A$_2$BX$_6$ vacancy ordered double perovskites, the B site has an oxidation state of 4+, and thus we use BX$_4$ as the binary phase. Most of the larger band gaps arise from double perovskites due to the lack of overlap between BX$_6$ octahedra owing to the alternation of B and B' cations in the structure. Decomposition energies tend to cluster near the 0 eV mark, and we usually consider a range of -0.5 eV to 0.5 eV as stable, keeping in mind theoretical uncertainties and possible entropic contributions. Other computational details can be found in past works \cite{htdft-1, htdft-2, htdft-3}. \\ 
%Note that this differs from the energy above the hull since we only consider binaries of constituent elements in their most stable phase.

\subsubsection{Materials Project Data}

\begin{figure}
    \centering
    \includegraphics[width=0.80\linewidth]{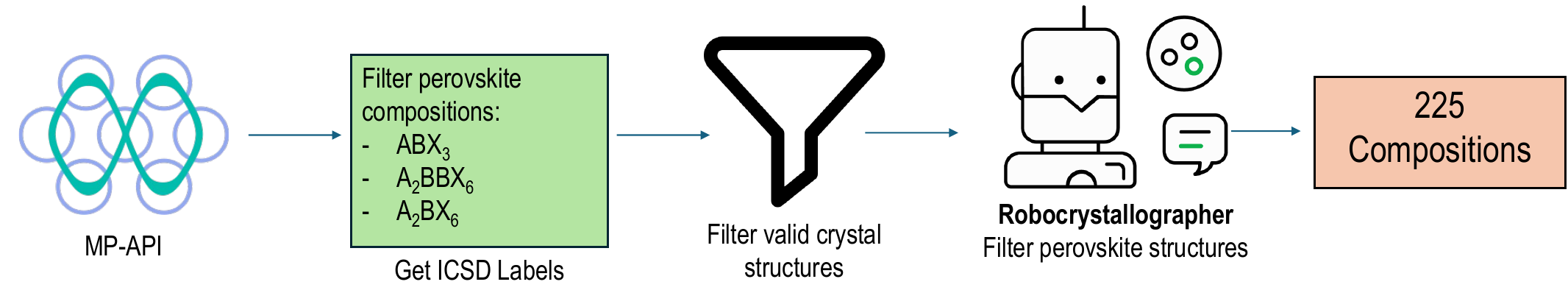}
    \caption{The workflow used to extract valid perovskite compounds from the MP \cite{mp} database using the Materials Project API}; this data is then combined with our DFT dataset following quick calculations using uniform computational parameters.
    \label{fig:mp-extraction}
\end{figure}

We also queried the Materials Project (MP) \cite{mp} database to find perovskite structures distinct from the compounds in our dataset and subsequently combined all the compounds into one dataset. We first filtered all possible compounds with general formulas of ABX$_3$, A$_2$BB'X$_6$, and A$_2$BX$_6$, where the anion X is either a halogen or chalcogen. The MP filtering workflow is pictured in \textbf{Figure \ref{fig:mp-extraction}}. This process led to a total of about 900 compounds. Then, we used the Robocrystallographer tool \cite{robocrystallographer} to filter out the compounds that did not contain specific keywords such as ``perovskite", ``perovskite-derived", and ``octahedral", following which a manual examination was performed to verify that the parsed compound is indeed a perovskite structure. This finally led to 225 compounds, out of which 111 were already present in the Mannodi group dataset. For the 114 unique compounds, we performed additional DFT computations using the exact same GGA-PBE input parameters and convergence criteria used in our dataset, thus obtaining their band gaps and decomposition energies. Subsequently, compounds with zero band gaps were excluded from the training set to ensure model stability, resulting in 58 new compounds which were added to our original dataset to yield an expanded dataset of 909 points. We also used the ``ICSD tag" (indicating experimental existence) available in the MP for any of the compounds to add synthesis labels to them. Out of the 225 compounds collected from MP, 33 had a valid ICSD tag, showing that they had been previously successfully synthesized. The computed properties pictured in \textbf{Figure \ref{fig:db}(b)} are for the combined PBE dataset of 909 unique compounds. \\

\subsection{Positive Unlabeled Learning}

To predict the synthesis likelihood of any material, one could utilize synthesis labels of known materials and derive similarity-based probability estimates. A label of ``1" can be assigned to compounds with clear evidence of laboratory synthesis (as reported in the literature), but a label of ``0" indicating clear non-synthesizability is not generally possible. For a given unknown compound, we can use the experimental positives and theoretical properties to estimate the likelihood of synthesis. To solve this task, in this work, we leveraged a semi-supervised machine learning technique called ``Positive-Unlabeled Learning" or PU learning, which has been used in recent years by multiple researchers to predict synthesizability of a variety of solid-state materials \cite{pu-perovs, mof-pu, pucgcnn, pu-walsh}. \\

\begin{figure}
    \centering
    \includegraphics[width=0.50\linewidth]{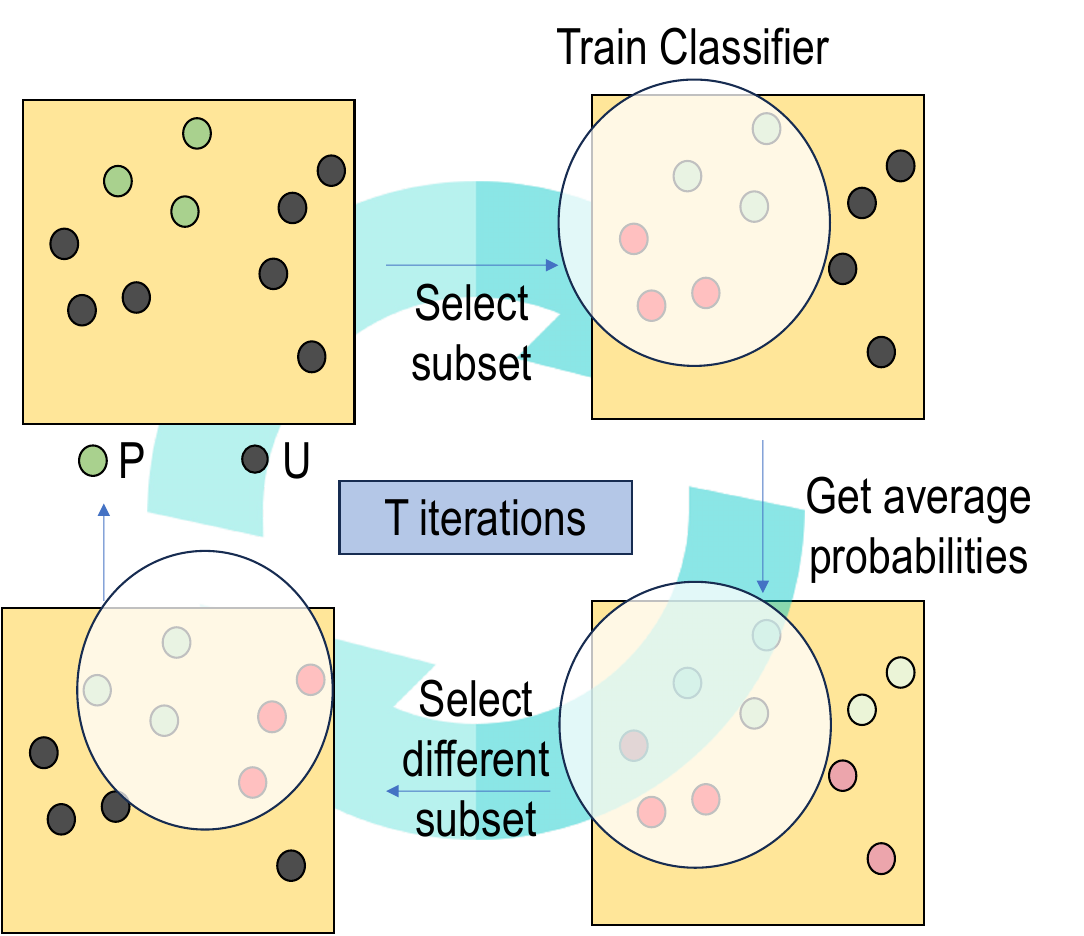}
    \caption{Transductive Bagging for PU Learning. Each subset selection uses a K-fold split performed R times. For each K-fold in R, the model is trained for T iterations. The figure has been adapted from \cite{cgcnn}.}
    \label{fig:tb}
\end{figure}

As the name suggests, PU learning works upon a dataset with many positively labeled points and many unlabeled points. A classifier model is trained on the dataset by randomly assigning negative labels to unlabeled points. By learning from the combination of known and arbitrary labels, the model assigns an ensemble-based probability to the unlabeled points. There are many learning paradigms within this framework, and we used the ``Transductive Bagging" approach introduced by Mordelet et al. \cite{pu-learning}. In this approach, the first training loop considers all the known positive samples (K) and an equal number of random unlabeled samples (U$_n$), which are assigned negative labels. The classifier learns the difference in their relations based on the input descriptors and assigns a probability to the remaining unlabeled points (U-U$_n$) based on what has been learned. This step is repeated for T iterations, and the probability is then averaged over all the unlabeled points across the T loops. This workflow is pictured in \textbf{Figure \ref{fig:tb}}. \\

As is evident from \textbf{Figure \ref{fig:db}(b)}, there is an imbalance in the DFT dataset in terms of compounds from different classes, since it is dominated by cubic ABX$_3$ HaPs. To remove this bias while sampling the data, we used a repeated K-fold sampling approach during PU learning. In each repetition R, we applied K-fold sampling and T bagging iterations for each fold. This leads to better representation of each unlabeled point. The same approach is also applied when choosing the test set for the best classifier. This is necessary because a fixed test set, as was used in previous studies on synthesizability predictions \cite{pu-perovs, pucgcnn, pu-walsh}, would introduce bias due to a class imbalance between types of perovskite compounds. ROC-AUC curves are eventually plotted to examine the training adequacy of the PU learning model. Since the positive label points are the only ones known with 100\% reliability, only the ``True Positive Rate" can be used as a valid metric on a sampled test set since all the unlabeled points are objectively unknown. We tested three different base-classifiers for training the PU learning models: Support Vector (SV), Random Forest (RF), and Decision Tree (DT). \\

\subsubsection{Descriptors.}

\begin{figure}
    \centering
    \includegraphics[width=\linewidth]{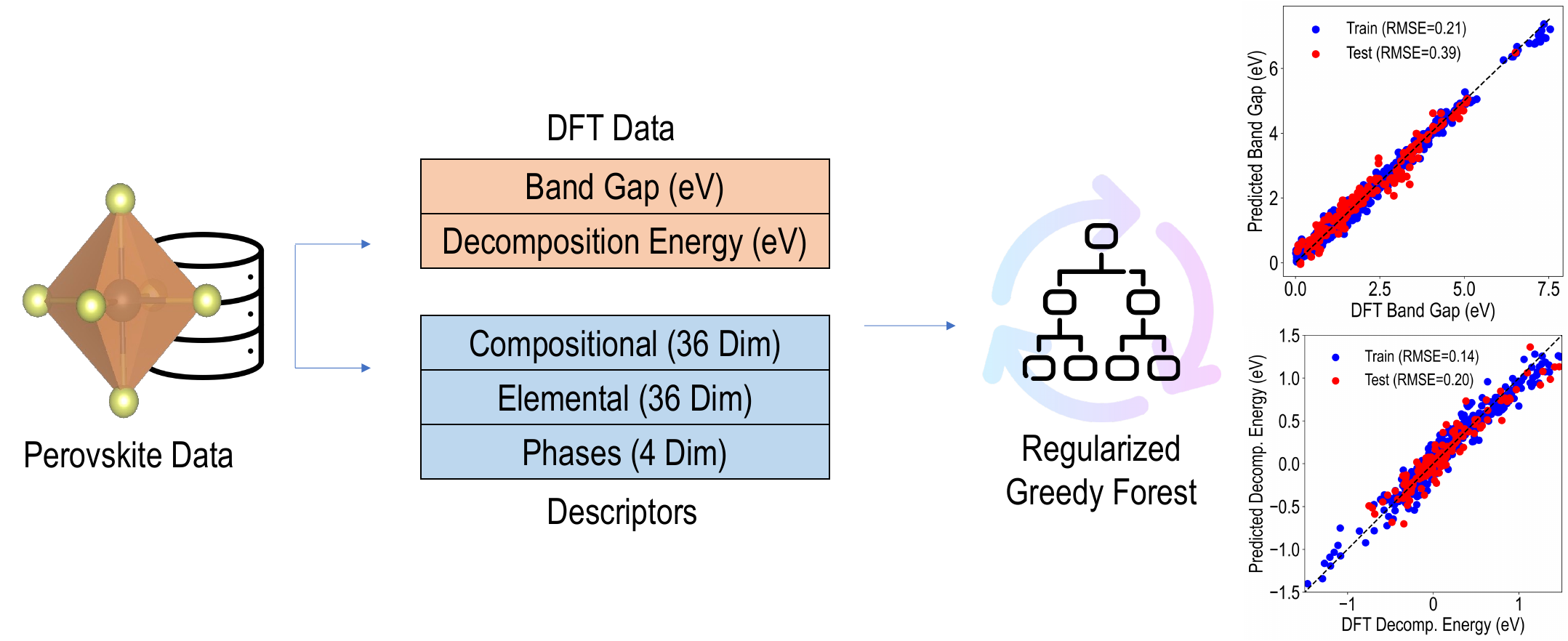}
    \caption{Regression models based on regularized greedy forests trained for the decomposition energy and band gap. These models can be used to predict the properties of any new perovskite based on its compositional and elemental descriptors.}
    \label{fig:rgf}
\end{figure}

Every perovskite compound is represented using an essential set of compositional inputs, elemental and molecular properties, and perovskite phase label descriptors, as pictured in \textbf{Figure \ref{fig:rgf}}. This is the same definition we used in multiple previous publications, which focused on ABX$_3$ HaP compounds. In our dataset, there are 36 distinct A/B/B'/X species, and their fractional occurrence in any compound is expressed using a 36-dimensional compositional vector. The composition vector is normalized by the formula unit to ensure it works uniformly across single and double perovskites. Well-known physical properties of all 36 species, such as ionic radii, electronegativity, and boiling point, are then used to create another 36-dimensional elemental properties vector, with 12 dimensions each for A, B/B', and X sites. These properties are weighted by the fractions of every species that occurs at any site in a given compound, such that this definition uniformly works for alloyed and unalloyed perovskites, and single and double perovskites alike. A four-dimensional vector is additionally used to show whether the compound has a cubic, tetragonal, orthorhombic, or hexagonal structure. \\ 

Elemental features such as electronegativity and boiling point provide the model with the chemical intuition for how the properties of constituent ions can affect the stability and synthesizability. On the other hand, the use of features such as ionic radii and the perovskite phase provides some structural information which helps assess phase stability better. The model could also benefit from specific synthesis conditions like temperature and pressure as well as information about likely synthesis routes and associated energy barriers, but the lack of consistent metadata about these aspects across the literature limits our ability to build a descriptor space from those quantities. Thus, the current descriptor choices only relate to the physical properties of the chemical species that make up any compound, and this would be sufficient for initial prediction and screening beyond which other synthetic information must be accounted for. \\

The resulting 76-dim descriptor (36 + 36 + 4) was used to first train regularized greedy forest (RGF) regression models to predict the perovskite decomposition energy and band gap. Parity plots shown in \textbf{Figure \ref{fig:rgf}} show the optimized models; the test prediction errors (shown using the root mean square error metric) reveal a roughly 95\% prediction accuracy. To train the synthesis classifier models, these descriptors were additionally fortified with 2 added dimensions for the decomposition energy and band gap. The purpose of doing this is to see if DFT information is useful in predicting the synthesis probability, rather than using only composition-based descriptors. The RGF model is used to predict the DFT properties of any novel compound. The complete set of 76 descriptors is presented in \textbf{Table S1}. \\

% \subsubsection{Decomposition Energy Labels}
% explain how we calculate decomposition energy and show the formula for the same

\subsubsection{Assigning Experimental Labels.}
The positive labels applied to compounds in our dataset are obtained from two sources: ICSD tags from the MP database as described above, and labels collected from across the published literature. The literature-based label extraction is crucial as updates on ICSD may lag behind publications for newly synthesized compositions, which can limit our positive labeled data. This is especially true for the general areas of halide and chalcogenide perovskites where the rate of new publications is substantial. Scanning through potentially thousands of peer-reviewed publications to obtain relevant perovskite compositions is an extremely time-consuming task. Research on halide and chalcogenide perovskites is thriving, and new publications emerge every day. Tools based on natural language processing (NLP) \cite{chemdataextractor, nlp-1, nlp-2, nlp-3} have frequently been used for targeted data collection from textual information. While such tools are efficient at gathering particular information, such as chemical formulas and properties, it is still challenging to determine whether and how compounds were synthesized in the laboratory. \\

\begin{figure}
    \centering
    \includegraphics[width=\linewidth]{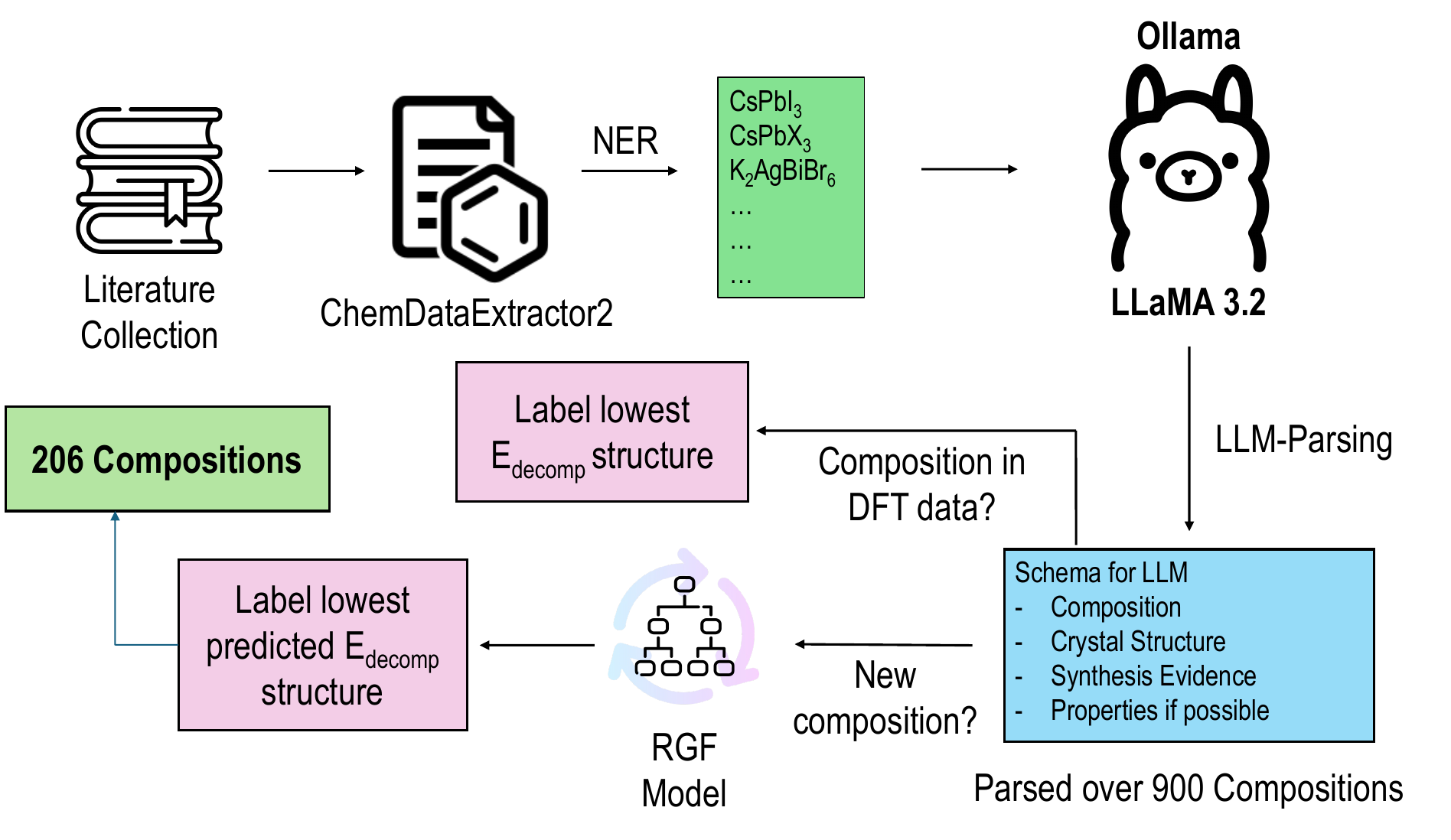}
    \caption{Our workflow for extracting experimental labels from the literature using the Llama3.2 model \cite{Llama32} supported via the Ollama \cite{Ollama} library. Meaningful new compositions thus obtained are assigned a perovskite phase based on the literature or based on our predicted decomposition energy values.}
    \label{fig:llm-extraction}
\end{figure}

State-of-the-art large language models (LLMs) are capable of efficiently extracting materials knowledge from large volumes of text \cite{llm}. Even so, processing entire PDFs of published papers using an LLM is time-consuming and exhausts many computational resources. Here, since our primary objective is to investigate the papers that report the synthesis of perovskite-type compositions, we only need to focus on the portions of text that list said compositions. This motivates our data collection pipeline pictured in \textbf{Figure \ref{fig:llm-extraction}}, where we extract chunks of text from a paper that contains a perovskite composition and pass them through an LLM. We used chain-of-thought inspired prompting, showing an example to the LLM to make it understand what we are looking for. For example, we directed the LLM to look for a composition that has an associated experimental method by giving it the names of the common characterization methods like X-ray diffraction (XRD), scanning electron microscope (SEM), etc.. The model is also shown keywords such as characterization patterns, deposition, experimental synthesis, etc., that can hint at the synthesis of a compound. \\

We first take the PDF of a publication and convert it into an XML format using docling \cite{Docling}, and then pass it through ChemDataExtractor2 \cite{chemdataextractorv2}, which uses Named Entity Recognition (NER) to filter portions that contain perovskite compositions. Here, the code tries to match the text to known chemical formulas of interest. We thus extract all the paragraphs that contain a perovskite-type chemical formula. The extracted data is then passed through an LLM using the Ollama \cite{Ollama} library that hosts multiple open-source local LLM schemes, providing a lightweight option to quickly screen through a large number of papers, and the output schema is handled at the end. If present, we extract the composition, crystal structure, evidence of synthesis, and properties of any compounds. Accurate collection of properties was found to be challenging and was discarded here in favor of simply obtaining synthesized perovskite compositions and phases. Using this pipeline, we processed over 1000 PDFs of perovskite literature collected by our group over the years and extracted more than 900 compositions. \\

We found that there were many duplicates among the collected set of compounds, and sometimes the compositions were expressed as, e.g., ``CsPbX$_3$", where X could be any halogen anion but was unspecified in the publication. It was also difficult to always parse the crystal structure information for the respective compounds, either due to model challenges or missing information. After cleaning and filtering the data, we obtained 206 unique compositions. Most of these compounds are alloys or contain different combinations of the previously discussed set of A/B/B'/X species. The relative distribution of cation and anion species in the positive and unlabeled sets is illustrated by the frequency charts in \textbf{Figure S6}, which reveal a similar distribution across both sets. This similarity ensures sufficient representation within the positive space, enabling reliable predictions of synthesizability for unlabeled compounds. We kept the compositional space fixed when extracting the compounds so as not to obtain a sparsely represented dimension (ionic species). Since the perovskite descriptor needs both composition and phase, the final step was to determine the preferred crystal structure for the compositions with missing phase information. To do this, we used the DFT-computed or RGF-predicted decomposition energy of a compound and assigned the lowest energy phase as the ground state. \\

The 206 compounds with positive synthesis labels extracted from the literature were added to the 33 compounds from MP with ICSD labels to yield a total positive label set of 239 compounds. From the DFT dataset of 909 compounds, 77 overlap with the experimental dataset, meaning that 832 compounds are synthetically unlabeled but have DFT properties. This additionally means 162 experimentally known compounds do not have DFT properties, and thus, the RGF models were used to obtain their DFT-accuracy decomposition energy and band gap. This combined dataset was then used to train the classifier models discussed next. \\

% The final dataset distribution is highlighted in \textbf{Table \ref{tab:distribution}}, which shows the unique datapoints with their original source. The dataset is classified into four categories based on data provenance. ``DFT" denotes compositions with properties derived from first principles calculations. ``Predicted" refers to compounds whose properties were generated by the RGF model. ``Experimental" means compounds with documented synthesis, and ``Theoretical" applies to those for which experimental evidence was not found.\\ 

% \begin{table}[]
% \begin{tabular}{|c|c|c|c|c|}
% \hline
% Source            & DFT & Predicted & Experimental & Theoretical \\ \hline
% Mannodi group     & 851 & 0         & 70           & 781         \\ \hline
% Materials project & 53  & 0         & 2            & 51          \\ \hline
% Literature        & 5   & 162       & 167          & 0           \\ \hline
% \end{tabular}
% \caption{A summary of the compiled dataset, detailing the number of compounds sourced from our group's calculations, the Materials Project, and the literature. Each entry is classified based on its data type: calculated from DFT, predicted by regression models, experimentally synthesized, or purely theoretical.}
% \label{tab:distribution}
% \end{table}

\begin{figure}
    \centering
    \includegraphics[width=\linewidth]{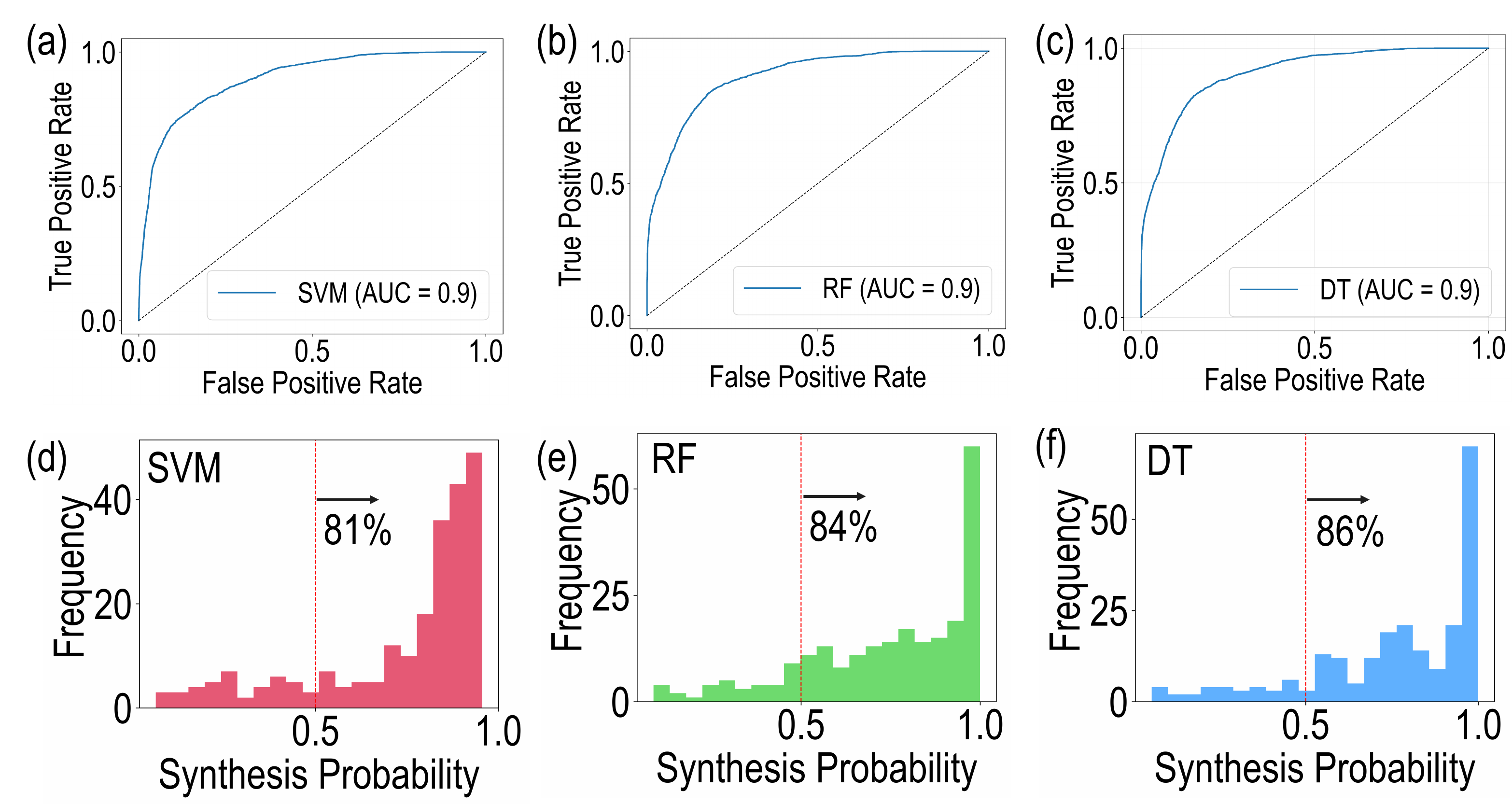}
    \caption{ROC-AUC curves for PU learning classification models trained using (a) Support Vector Machines (SVM), (b) Random Forest (RF), and (c) Decision Trees (DT). Bar charts in (d), (e), and (f) show predicted probabilities for the test set, with the DT model achieving the best True Positive Rate (TPR) of 0.86.}
    \label{fig:tpr_auc_comparison}
\end{figure}

\section{Results and Discussions}

We tested three types of classifier models: a Support Vector (SV) classifier, a Random Forest (RF) classifier, and a Decision Tree (DT) classifier, each as implemented within the Scikit-learn library. To select the best base classifier, we rigorously trained three individual models and plotted their ROC curves along with the predicted probabilities for the test set. These models were trained on the combined dataset of unlabeled compounds (832 points) and compounds with experimental labels (239 points), within a probabilistic framework where (experimentally) unlabeled points were assigned random positive or negative labels through the ensemble learning process, as explained earlier. We found that each model performed significantly better than a purely random classifier, as pictured in \textbf{Figure \ref{fig:tpr_auc_comparison} (a), (b), (c)}. The DT classifier leads to the best performance with an area under the curve (AUC) of 0.908. This success can also be gleaned from the high true positive rate (TPR) of 86\% for the DT model on the cumulative random test set, pictured in \textbf{Figure \ref{fig:tpr_auc_comparison} (f)}. In comparison, SV and RF models show TPR values of 81\% and 84\%, respectively. Notably, some data points extracted from the literature populate low-probability regions. This is a useful outcome, as it indicates the model may be flagging false positives introduced by the automated extraction process. Thus, we used the DT classifier model going forward to make any new predictions. Optimum values of R = 10, K = 5, and T = 500 were found to yield the best AUC and TPR values for the DT model. Performance comparisons between different model parameters are provided in \textbf{Figure S2}. \\

\begin{figure}
    \centering
    \includegraphics[width=0.8\linewidth]{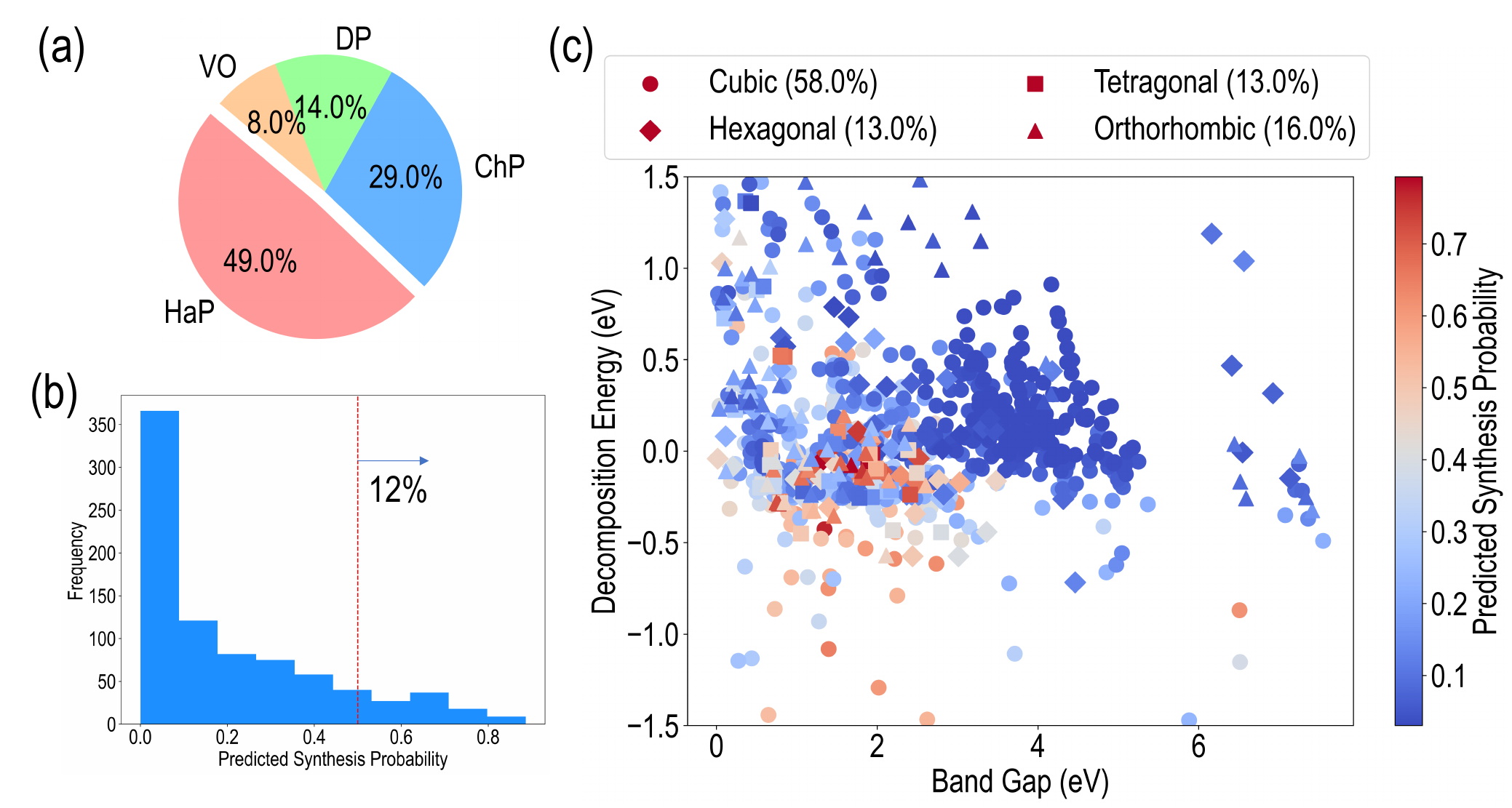}
    \caption{(a) Distribution of different types of perovskites predicted to be synthesizable. (b) Predicted synthesis probabilities for the unlabeled set of compounds. (c) DFT-computed decomposition energy plotted against the band gap for unlabeled compounds, with the color chart showing predicted synthesis probabilities. The crystal structure legend shows what \% of different perovskite phases are predicted to have synthesizability $>$ 50\%.}
    \label{fig:final_result}
\end{figure}

We used the optimized DT classifier model to predict the synthesis probabilities for all the unlabeled samples, and the results are presented in \textbf{Figure \ref{fig:final_result}}. 100 out of the 832 unlabeled compounds were predicted to have a synthesizability $>$ 0.5. A distribution of the different types of perovskites across these 100 compounds is shown in \textbf{Figure \ref{fig:final_result}(a)}, and a histogram of the predicted synthesizabilities across the dataset is pictured in \textbf{Figure \ref{fig:final_result}(b)}, showing that only a small fraction (12\%) of the compounds have a greater than 50\% likelihood of being made in the lab. \textbf{Figure \ref{fig:final_result}(c)} shows the entire DFT dataset as a decomposition energy vs band gap plot like before, with the scatter points colored by their predicted synthesis probabilities. From the \textbf{Figure \ref{fig:final_result}(a)}, we can contend that our prediction spans across different perovskite types and crystal structures, even with limited positive labels, showing the robustness of this approach. \\

We find that a vast majority of the highly synthesizable compounds (darker red points) fall in the decomposition energy $<$ 0.5 eV and band gap less than 3 eV range. This shows a good correlation between decomposition energies and synthesis likelihood, which is also evident from the correlation values in \textbf{Figure S3}. Many of these compounds are thus ideal candidates for single- or multi-junction solar absorption applications. A majority of the synthesizable compounds are also cubic, which is a consequence of the imbalance in the dataset, and especially the fact that all double perovskites (DP and VO compounds) are only simulated in the cubic phase. From \textbf{Figure \ref{fig:final_result}(a)}, it can be seen that 49\% of the most synthesizable compounds are halide perovskites, 29\% are chalcogenides, and only around 20\% are double perovskites. This is a consequence of the dataset demographics, but it should be noted that potentially thousands of new synthesizable compounds could be designed within each of these sub-spaces by enumerating new compositions and making on-demand predictions. \\

We extracted feature importance values from the DT classifier model, and results for the most important features are pictured in \textbf{Figure S3}. The atomic weight of the B-site cations shows by far the highest importance, followed by other physical properties of A and B cations. This is an expected result as the thermodynamic stability of a perovskite is dictated by the arrangement of constituent BX$_6$ octahedra. While the DFT-computed decomposition energy and band gap are comparatively less important, they do feature in the top 10 most important dimensions, thus showing the importance of using computed properties for improving synthesis predictions. The ionic radii and weight/size of A/B/X ions are also important since they contribute to the tolerance and octahedral factors, well-established perovskite formability criteria. Features such as electronegativity and heat of fusion are known to correlate with general electronic and energetic behavior and thus make sense as contributors to the synthesizability. \\

\textbf{Table \ref{tab:top2}} shows the top 2 most synthesizable compounds predicted for each perovskite type (HaP, ChP, DP, VO). All compounds are cubic, have generally low decomposition energies, and photovoltaic-suitable band gaps with a couple of exceptions. There is a good variation in the types of cations observed in these compounds. There is a prevalence of Cs/FA and I/Br alloying among the most synthesizable HaP compounds, which has been shown in the past as a route to stabilize perovskites and reduce phase separation. For ChP ABX$_3$ compounds, Sr is prevalent at the A site, and the compounds are generally sulfides or selenides. This is consistent with the recent thrust from researchers to focus on Sr-based perovskite chalcogenides \cite{sr-perovskites} to move away from Ba-based compositions that suffer from larger than desired electron concentrations \cite{defect_bazrs3}. One of the vacancy-ordered double perovskites identified by the model is the compound Cs$_2$PtBr$_6$  \cite{cs2ptbr6}, which is actually known to have been synthesized and was not included in our training dataset. \\ 

One of the main caveats about our synthesis prediction models is that it is almost definitely missing a notable number of compounds with positive labels that exist in the literature but have not been captured by our knowledge extraction framework, and adding all this data to the training set will certainly improve prediction accuracy and applicability. Furthermore, it is known that many halide and chalcogenide perovskites could exist in multiple phases at different conditions of temperature and pressure, while we currently only consider the single most stable phase at ambient conditions. It stands to reason that we could easily generate descriptors for compounds in multiple phases and obtain their synthesis probability values, which may reveal unexpected synthesizable phases for given compositions. \textbf{Table \ref{tab:top2}} also lists some compounds such as Cs$_2$ScAgBr$_6$ and Cs$_2$HfI$_6$ which have not been previously synthesized (as far as we have seen in the literature) but are similar to other compositions that have indeed been made. These compounds are already part of our dataset and have been optimized using GGA-PBE. Their relaxed structures are uploaded as part of SI and some of them are visualized in \textbf{Figure S7}. A complete list of the 100 compounds with DFT properties that were predicted to have a synthesis probability of $>$ 50\% is provided in the Supplementary Information. \\

\begin{table}[]
\centering
% This command ensures the multi-line header is centered
\renewcommand\theadalign{bc}
\begin{tabular}{|c|c|c|c|}
\hline
\textbf{Composition} & \textbf{Band Gap (eV)} & \thead{\textbf{Decomposition}\\\textbf{Energy}\\\textbf{(eV p.f.u.)}} & \textbf{Synthesis Probability} \\ \hline
Cs$_{0.25}$FA$_{0.75}$PbI$_3$ & 1.71 & 0.03 & 0.89 \\ \hline
MAPbBr$_{0.5}$I$_{2.5}$ & 1.79 & 0.03 & 0.85 \\ \hline
Cs$_2$PtBr$_6$ & 1.31 & -1.08 & 0.74 \\ \hline
Cs$_2$ScAgBr$_6$ & 2.74 & -0.62 & 0.69 \\ \hline
Cs$_2$HfI$_6$ & 2.21 & -0.59 & 0.69 \\ \hline
Cs$_2$AgSbCl$_6$ & 1.39 & -0.75 & 0.68 \\ \hline
SrZrS$_3$ & 1.44 & 0.53 & 0.65 \\ \hline
SrHfS$_3$ & 1.63 & 0.53 & 0.61 \\ \hline
\end{tabular}
\caption{Examples of the most synthesizable compounds in the DFT dataset, along with their PBE properties and synthesis probability values.}
\label{tab:top2}
\end{table}

% In our previous work, we hypothesized five compounds that would be viable for water-splitting\cite{ml-amk-1}, which were also hypothetical compositions. We also test out those compositions using our model and predict their synthesizability.[need to add this, should be good]\\

Finally, our regression models for DFT property prediction and the DT classifier model for synthesizability prediction are ready to be applied across the combinatorial space of all possible perovskite halides and chalcogenides within the defined chemical space of A/B/B'/X ions. To keep the problem tractable and for purposes of demonstration, we used only ABX$_3$ compounds here. We enumerated a total of 20,000 novel compositions across different phases by applying constraints to the cation and anion choices to keep them within the chemical space we described above. We also ensure the compounds already existing in the dataset are not enumerated. Each composition enumerated is then repeated in the four different phases. Different cation and anion species were fit into the canonical ABX$_3$ formula in appropriate oxidation states and mixing fractions to ensure overall charge neutrality. The generated compounds include unalloyed compositions as well as X-site alloys with only halogen anions, only chalcogen anions, and both halogen and chalcogen anions. The chalco-halide perovskites are included to demonstrate that our model can be utilized even for these cross-species materials, and it has been reported recently that chalco-halide compounds show desirable stability and defect tolerance \cite{chalcohalide}. Descriptors were generated for all 20,000 compounds, the RGF models were used to predict their (GGA-PBE accuracy) decomposition energy and band gap, and then the DFT properties were added to the descriptor list and the DT classifier model was used to predict their synthesis probabilities.\\ 

\begin{figure}
    \centering
    \includegraphics[width=0.65\linewidth]{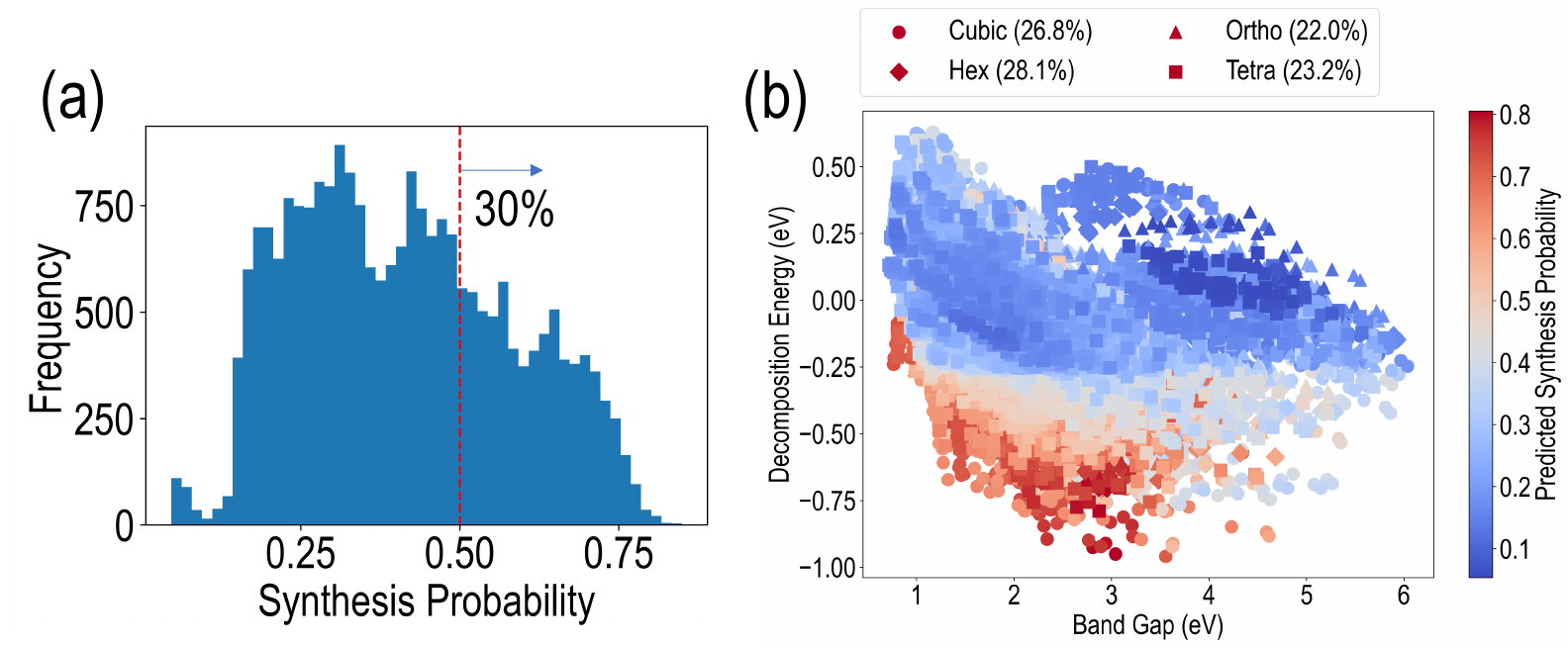}
    \caption{(a) Distribution of predicted synthesis probabilities across the set of 20,000 novel enumerated compounds. (b) ML-predicted decomposition energy plotted against the band gap across the entire set of enumerated compounds, with the color bar showing predicted synthesis probabilities. The crystal structure legend shows what \% of different perovskite phases are predicted to have synthesizability $>$ 50\%.}
    \label{fig:hypothetical}
\end{figure}

\begin{table}[]
\centering
% This command ensures the multi-line header is centered
\renewcommand\theadalign{bc}
\begin{tabular}{|c|c|c|c|c|}
\hline
\textbf{Composition} & \textbf{Phase} & \textbf{Band Gap (eV)} & \thead{\textbf{Decomposition}\\\textbf{Energy}\\\textbf{(eV p.f.u.)}} & \textbf{Synthesis Probability} \\ \hline
CsZrFS$_2$ & Cubic & 3.04 & -0.95 & 0.85 \\ \hline
CsZrClS$_2$ & Cubic & 2.81 & -0.93 & 0.84 \\ \hline
CsZrFSSe & Cubic & 2.94 & -0.91 & 0.84 \\ \hline
CsPbClBr$_2$ & Cubic & 2.15 & -0.21 & 0.84 \\ \hline
CsPbCl$_2$Br & Cubic & 2.32 & -0.22 & 0.83 \\ \hline
CsZrFS$_2$ & Hex & 2.95 & -0.71 & 0.83 \\ \hline
CsZrClS$_2$ & Hex & 2.74 & -0.69 & 0.83 \\ \hline
CsZrFSSe & Hex & 2.87 & -0.70 & 0.82 \\ \hline
CsAgF$_2$Br & Cubic & 2.33 & -0.40 & 0.82 \\ \hline
CsHfFS$_2$ & Cubic & 3.21 & -0.84 & 0.81 \\ \hline
\end{tabular}
\caption{A list of novel compounds with high predicted synthesis probability. Also shown are their regression model-predicted (PBE accuracy) properties.}
\label{tab:hypothetical}
\end{table}

\textbf{Figure \ref{fig:hypothetical}} shows histograms of the distribution of the predicted synthesizability across the space of the enumerated compounds, and a plot between the predicted decomposition energy and band gap of all these compounds, with scatter points colored by the predicted synthesis scores. We find that around 30\% of the compounds have a synthesis probability $>$ 0.50, but the number of compounds drastically reduces for the highest probability values. Only 13\% of the compounds show a value greater than 0.65. From \textbf{Figure \ref{fig:hypothetical}(b)}, it is clear that there is a correlation between the synthesis probability and the decomposition energy: nearly all the most synthesizable compounds have negative decomposition energies, and there is a noticable blue to red gradient from positive to negative decomposition energy values, similar to the original DFT dataset. This plot reveals hundreds of potential compositions with PV-suitable band gaps and high resistance to decomposition + synthesizability. Some of the top predicted compounds are highlighted in \textbf{Table \ref{tab:hypothetical}}, with most of them adopting the cubic phase. Multiple compounds here are chalco-halides while the rest involve halogen mixing. The compound CsPbCl$_2$Br has actually been reported experimentally before \cite{cspbcl2br}, thus providing some additional validation for our predictions. These models thus enable the study of various alloy compositions without the need for computationally expensive DFT calculations, allowing for the derivation of composition-property correlations. As an example, \textbf{Figure S8} shows heatmaps of the RGF-predicted band gaps and predicted synthesis scores for Ba(Zr,Hf,Ti)S$_3$ orthorhombic phase alloys, revealing some important information about how the incorporation of Zr/Hf/Ti at the B-site influences the stability and properties of interest. The predicted properties and synthesis probabilities of all 20,000 compounds are also provided in the SI. \\

\section{Future Outlook}

\begin{figure}
    \centering
    \includegraphics[width=\linewidth]{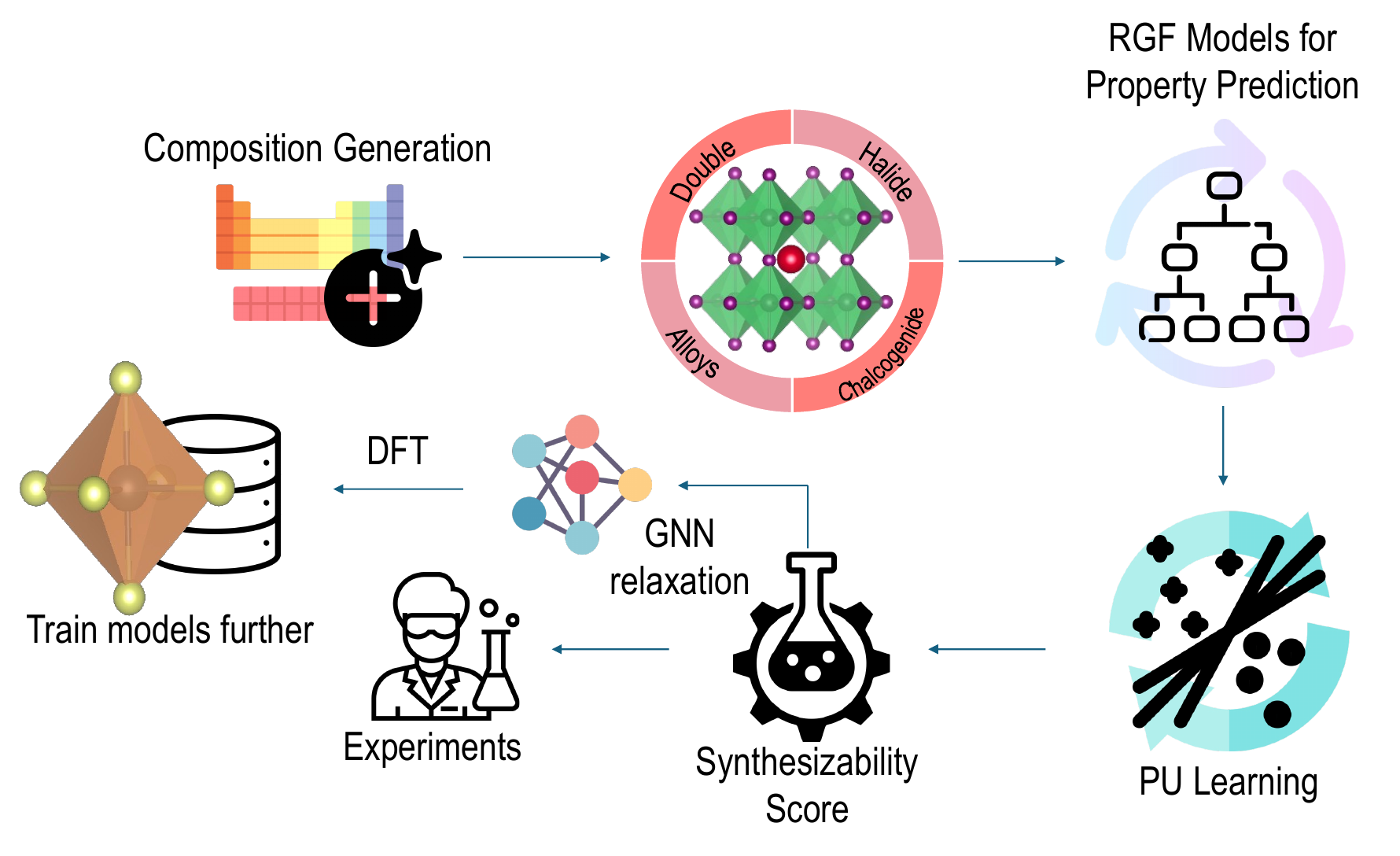}
    \caption{The proposed workflow for discovery of novel perovskites, combining regression models, synthesis predictions, GNN-accelerated DFT, and final computations and experiments for validation.} 
    \label{fig:final_workflow}
\end{figure}

\textbf{Figure \ref{fig:final_workflow}} shows our proposed framework for computational discovery of novel perovskites, powered by the models developed in this work. This high-throughput screening workflow begins by identifying all possible charge-neutral compositions that satisfy a single or double perovskite formula. Initial screening will be performed using RGF models to predict the DFT-accuracy decomposition energy and band gap, which serve as surrogates for material stability and electronic properties. This stage is flexible, allowing the models to be tuned for derived, application-specific properties to further refine the candidate pool. The resulting set of filtered compounds will then be assessed for their synthesizability using the PU learning classifier model developed in this study. Candidates deemed to have a high synthesis probability will be assigned to template crystal structures; that is, structural representations will be created for them using existing prototype phases or structures of known similar materials. At this point, these structures must be optimized at DFT accuracy, following which other calculations can be performed to sample various properties of interest: optical absorption, defect energetics, phonon dispersion, etc. \\

Given the expense of running DFT simulations at GGA accuracy or higher-level functionals, especially using large supercells to simulate alloys or defects, it is vital to implement ML-driven acceleration. For instance, DFT geometry optimization can be augmented with graph neural network (GNN)-based interatomic potentials \cite{m3gnet, mace-off}, such as the models recently developed in our group \cite{perovsiap}. GNN+DFT will be used to perform efficient structural relaxation to obtain ground state configurations. Interatomic potentials can further aid in navigating the defect landscape and performing perturbation-based calculations. To accelerate our DFT workflows, we aim to develop a unified model by employing a fine-tuning strategy analogous to that of the current Perovs-IAP framework. This new model will be trained on a comprehensive dataset encompassing all types of perovskite structures. The final stage would involve performing rigorous high-accuracy DFT computations exclusively for the curated subset of best materials. Additionally, these compounds will be recommended to experimental collaborators for synthesis and characterization; any new experimental data will serve to validate the predicted synthesizability and computed properties. This multi-stage computational screening approach is designed to dramatically reduce the number of required calculations, thereby accelerating both the simulations and the discovery of novel perovskite materials. \\

We envision a variety of extensions and improvements to this work in the near future:
\begin{itemize}
    \item An expanded DFT dataset that includes compounds from lesser-represented chemical subspaces, such as chalco-halide perovskites.
    \item Extraction of more experimentally known compounds from the literature will continue, especially using new publications that emerge every day.
    \item Extension to other DFT functionals, including hybrid HSE06 calculations with and without spin-orbit coupling \cite{soc} to improve electronic band edges and gaps; this will lead to multi-fidelity models combining GGA and HSE06 data.
    \item An expanded chemical space which includes a series of other possible A/B/B' cations, such as 3d, 4d, and 5d transition metals, and a variety of other organic molecular cations \cite{mol-1, mol-2, dmetal}.
    \item The LLM extraction pipeline can be enhanced by using better prompts and experimenting with different models. 
    \item Approaches other than PU learning can be explored to make synthesis probability predictions \cite{syncotrain}; additionally, the synthesizability framework can be integrated with structure-based GNN models to eliminate the need for hand-crafted descriptors, especially as the dataset swells in size.
    \item Extensions can be made to perovskite derivatives such as A$_3$B$_2$X$_9$ compounds, layered perovskites, and lower-dimensional (2D, 1D, 0D) perovskites, as well as to other properties of interest from DFT. 
    \item Finally and very importantly, it is vital that our predictions lead to experimental validation and discovery of new perovskites. Efforts are currently underway for synthesizing some of the compounds predicted to have the highest synthesis probabilities, and we welcome interest from the community in taking our predictions to the laboratory.
\end{itemize}

% In the future, we will bolster our model by running additional DFTs and collecting systematic experimental data from the literature. We would enhance our LLM extract pipelines by using better prompts and experimenting with different models. The elemental space would be expanded to cover the entirety of the periodic table, with all the different types of perovskites included. Our aim is to create a uniformly represented dataset among the different types of perovskite to make our predictive models better. This would include both expanding the chemical space and including chalco-halide perovskites. This dataset would then become a foundational basis for our data-driven workflow shown in the \textbf{Figure \ref{fig:final_workflow}}. \\

\section{Conclusions}
In this work, we developed a framework to predict properties at DFT accuracy along with the synthesis probability of perovskites by combining descriptor-based regression models and the Positive-Unlabeled (PU) learning approach. The models were trained on a chemically and structurally diverse dataset encompassing halide and chalcogenide single perovskites adopting cubic, orthorhombic, tetragonal, or hexagonal phases, as well as cubic vacancy-ordered and regular double perovskites. Experimental labels (indicating synthesis success) were assigned to compounds using ICSD tags in addition to information extracted from the scientific literature using large language models. The PU learning synthesis prediction model based on a decision tree classifier showed a true positive rate of 0.86 and an area under the curve of $>$ 0.90 on a held-out test set. This model was first used to predict the synthesizability of the entire DFT dataset of 909 compounds, followed by an expanded and enumerated set of 20,000 compounds which included regression-based decomposition energy and band gap predictions. Hundreds of promising stable and synthesizable compounds with potentially attractive optoelectronic properties are thus identified. This framework can be used in perpetuity for accelerated perovskite discovery. \\

% The descriptors included both fundamental elemental properties and computed properties at the PBE level of theory. Our model achieves a True Positive Rate (TPR) of 0.84 and an Area Under the Curve (AUC) of 0.9 on a held-out test set; notably, we show that a model based solely on composition can also yield strong predictive performance. When applied to 832 unlabeled compositions, our model identified 100 as potentially synthesizable, at least half of which represent novel, un-synthesized materials. Furthermore, feature importance scores indicate that the B-site element identity and the decomposition energy are critical drivers of synthesizability. This model, by identifying approximately 100 novel and promising compositions, provides clear guidance within the vast chemical space and will be integrated into our perovskite discovery pipeline to aid experimentalists in prioritizing synthesis targets.

\section{Acknowledgments}
A.M.K. acknowledges support from the Purdue University School of Materials Engineering. This research used resources hosted by the Rosen Center for Advanced Computing (RCAC) clusters at Purdue.

\section{Data Availability}
Datasets with DFT and experiment labels, code for LLM-based data extraction, and code for regression and classifier models are made available with the Supplementary Information. These include:
\begin{itemize}
    \item CSV files of the training data, experimental labels, and predicted synthesizability of the unlabeled DFT data and enumerated compounds.
    \item CSV file containing DOIs of all the papers used to extract the information using the LLMs.
    \item A Jupyter Notebook to run the LLM model and extract compositions that have been synthesized.
\end{itemize}
Data and code can also be accessed using our nanoHUB tool \cite{nh-tool}.

\clearpage
\section{References}
\bibliographystyle{chem-acs}
\bibliography{references}
\end{document}